\documentstyle[aps,float,epsfig]{revtex}  

\twocolumn

\begin{document}
\draft
\preprint{HEP/123-qed}
\wideabs{
\title{\Large\bf Searching for vortex structures in high Reynolds number
turbulence}
\author{\normalsize{S. I. Vainshtein}\\
{\small\it Department of Astronomy and Astrophysics, University
of Chicago, Chicago, Illinois 60637}}
\date{\today}
\maketitle

\begin{abstract}
\normalsize{
\noindent
In experimental  study of very high Reynolds number turbulence, we found 
evidences
that there are distinguished vortex structures in the intermediate range, that
is, between the Kolmogorov and Taylor microscales, where they are indeed
expected to be present. These structures are responsible for the intermittency,
and, in the same time, they contribute into asymmetry of turbulent
statistics, the latter following from the Kolmogorov law. 
}
\end{abstract}
}

\narrowtext

\section{Introduction}
\label{1}

The presence of organized structures in fully developed turbulence was
demonstrated recently both in physical experiments, and in numerical
simulations. Most of the studies are devoted to vortex tubes (or
``worms", or`` sinews"), see, e.g., \cite{Moffatt}--\cite{Bob}.

Generally, these organized structures may be considered at the same time
as a source (or manifestation) of intermittency, because normally they
correspond to a small dimension (of the size of Taylor microscale, or 
Kolmogorov 
microscale), although they are stretched in the other directions over  a
``macroscopic" length. 

One remarkable feature of these vortex structures is that they present not only
the intermittency of turbulence, but also contribute into the asymmetry of 
the PDF's \cite{1},\cite{2}. As was
suggested in \cite{94},  the asymmetric statistics of the turbulence is indeed
related to the intermittency. This connection was summarized in the so-called
ramp-model, the latter being only empirical. The considerations of the 
formation of vortex
structures, when the intermittency and asymmetry appear simultaneously,
have resulted in the development of the ramp-model into a more sophisticated 
version which is now called the bump-model.

The data used in this paper are
based on Taylor's hypothesis, and therefore do not provide direct measurements
of vortices. However, these vortices should be manifested in different ways, such
as appearance of intermittency, etc.  
 We provide here 
evidences that these vortices are present in fully developed turbulence. We
believe that the
disadvantages of necessary indirect studies of vortices presented in this paper,
are outweighed by the fact that we are studying a very high Reynolds turbulence.

The paper is organized as follows. Section ~\ref{2} is still introductory,
presenting (mostly known)
notions of turbulence statistics, related to the vortices. The main purpose of
this section is to explain, what can we expect to observe from experimental data
if the vortices are present, and also to
 introduce denotations we will use in the rest of the paper;
and thus some repetition of previous results seems to be inevitable. On the
other hand, the bump-model is introduced in Sec. \ref{2b}.
Section
~\ref{3} starts with suggestions how to improve the two extreme cases, or
models, considered in Sec. ~\ref{2}. The answer is in the middle: an
intermediate model is suggested. The rest of the section is devoted to
comparison of this  model with the experimental data, namely, with 
the observed structure  functions of
low orders. Section ~\ref{4} presents evidences
of existing vortices. Namely, it is shown that the intermittency is
substantially enhanced in the intermediate range: from studying the flatness
of the velocity increments statistics,
Sec. ~\ref{4a}, and from the box counting, Sec. ~\ref{4b}. An extensive
discussion of both extreme case models, and also of the intermediate model is
given in Sec. ~\ref{5}. It is argued that this model may give an insight into
the question of the intermittency formation, which proves to be intimately
connected with the asymmetry of turbulence. Finally, concluding remarks are
presented in Conclusion, Sec. ~\ref{6}.

\section{ Basic ideas and denotation }
\label{2}

We consider structure functions,
\begin{equation}
{\cal{S}}_n(r)=\langle (\Delta_r u)^n \rangle,
\label{structure}
\end{equation}
where $\Delta_r u=u(x+r)-u(x)$, and $n$ is an integer,
 and generalized structure functions,
\begin{equation}
S_q(r)=\langle |\Delta_r u|^q \rangle,
\label{structure1}
\end{equation}
where $q>0$.

We mention an exact result, obtained directly from Navier-Stokes equation,
\begin{equation}
{\cal{S}}_3(r)=-\frac{4}{5}\varepsilon r +6\nu \frac{\partial}{\partial r}
S_2(r),
\label{K}
\end{equation}
\cite{law}, 
and in inertial range where the viscosity term can be neglected, we recover the
4/5-Kolmogorov law,
\begin{equation}
{\cal{S}}_3(r)=-\frac{4}{5}\varepsilon r.
\label{K1}
\end{equation}

Data are from atmospheric turbulence measurements, about 35 meters  
above the ground, obtained by hot-wire anemometer, mean wind speed  
was 7.6 m/sec, root-mean-square velocity 1.3 m/sec. Using Taylor's  
hypothesis and local isotropy assumption to obtain dissipation, one  
obtains the Taylor microscale Reynolds number to be 9540 and the  
Kolmogorov microscale $\eta$ to be 0.57 mm. The data were sampled at 5 kHz  
and the file consisted of 10 million data points (courtesy of Sreenivasan).
The data  are treated in the spirit of the Taylor
hypothesis, that is, the time series is treated as one-dimensional cut of
the process (for more detail, see \cite{similarity,98,2000}).

\subsection{ No intermittency. Asymmetry in K41}
\label{2a}

Consider classical picture of turbulence due to K41 \cite{k41}. 
That is,
the turbulence is presented with an ensemble of multiscale cells, from integral
scale $\ell$ to smallest Kolmogorov scale $\eta \sim\ell/R^{3/4}$,
where $R\approx v\ell/\nu$, the Reynolds number, and 
$v=\sqrt{\langle v_x^2+v_y^2
+v_z^2\rangle}$, the mean square velocity,
with $|\Delta_r u|\sim r^{1/3}$. 
Then, in inertial range, i.e., at $\eta< r<\ell$,
\begin{equation}
S_q(r)\sim r^{q/3}.
\label{K41}
\end{equation}

We define the Taylor microscale,
\begin{equation}
\lambda=\sqrt{\frac{\langle u^2\rangle}{\langle (\partial_x u)^2
\rangle}}, 
\label{Taylor}
\end{equation}
see, e.g., \cite{book}. Then,
\begin{equation}
\lambda\sim\frac{\ell}{R^{1/2}}, ~~~{\rm and}~~~
R_\lambda=\frac{\langle u^2\rangle^{1/2}\lambda}{\nu}\sim R^{1/2}.
\label{Taylor1}
\end{equation}
It follows from (\ref{Taylor}) that
\begin{equation}
\lambda =15^{1/4}\eta R_\lambda^{1/2}, 
\label{Taylor2}
\end{equation}
see, e.g., \cite{text}, formula (3.2.18),  we  use this expression  below.

Let us proceed to the asymmetry, dictated by the Kolmogorov law (\ref{K1}).
Denote  $\lambda_{1,2,3}$ -- three eigen values of the symmetric part of the
matrix $\partial_j v_i$, and suppose that
\begin{equation}
\lambda_1\le \lambda_2\le \lambda_3. 
\label{lambda}
\end{equation}
It is known \cite{Townsend}, that the quantity defining skewness
of $\partial_x u$ field, namely $\langle
(\partial_x u)^3\rangle$, is 
$=(8/35) \langle\lambda_1\lambda_2\lambda_3\rangle$, and
therefore we calculate  $\langle\lambda_1\lambda_2\lambda_3
\rangle$. 
As $\lambda_1+\lambda_2+\lambda_3=0$ (due to the
incompressibility), (\ref{lambda}) can be replaced by
\begin{equation}
-\frac{\lambda_3}{2}\le\lambda_2\le \lambda_3.
\label{lambda1}
\end{equation}
We thus consider conditional mean value ($\lambda_3$ is fixed)
$$
\langle \lambda_1\lambda_2\lambda_3|\lambda_3\rangle=
-\langle (\lambda_2+\lambda_3)\lambda_2\lambda_3\rangle=$$
\begin{equation}
\int_{-\lambda_3/2}^{\lambda_3}
p(\lambda_2|\lambda_3)[-(\lambda_2+\lambda_3)\lambda_2\lambda_3]d\lambda_2,
\label{e3}
\end{equation}
where $p(\lambda_2|\lambda_3)$ is  conditional 
probability distribution of $\lambda_2$.
 The quantity $\lambda_1\lambda_2\lambda_3$ 
as a function of $\lambda_2$   is highly asymmetric and 
mostly  negative in the range (\ref{lambda1}). If, for example,
we suppose that $p(\lambda_2|\lambda_3)$ is an uniform distribution, i.e.,
$$p(\lambda_2|\lambda_3)=\frac{1}{(3/2 )\lambda_3 },$$
which at first sight seems to be  quite a realistic assumption, then, according to
(\ref{e3}), $\langle \lambda_1\lambda_2\lambda_3|\lambda_3\rangle=
-\lambda_3^3/2<0$. In fact, the contribution of the negative part of
$\lambda_1\lambda_2\lambda_3$ is 10 times larger than the positive part; 
that is,  in order to make the integral in (\ref{e3}) positive, the
distribution $p(\lambda_2|\lambda_3)$ should be highly asymmetric. Namely,
$p(\lambda_2|\lambda_3)$ should be more than 10 times larger in the negative range of
(\ref{lambda1}) than in the positive.
As this kind of  assumption concerning the PDF $p(\lambda_2|\lambda_3)$ does 
not seem to be  reasonable,
it follows from these
considerations that $\langle
(\partial_x u)^3\rangle$, and therefore the skewness should be negative.

In a more natural way, one may consider
the strain parameter,
$$
{\tilde{\lambda}}=\frac{\lambda_1-\lambda_2}{\lambda_1+\lambda_2}=\frac{
2\lambda_2+\lambda_3}{\lambda_3},
$$
introduced by Moffatt et al. \cite{Moffatt1}, and, according to
(\ref{lambda1}), $0\le {\tilde{\lambda}}\le 3$. Then,
$$
\lambda_1\lambda_2\lambda_3
=\frac{\lambda_3^3}{4}(1-{\tilde{\lambda}}^2).
$$
The interval where this expression is positive (i.e., $0\le {\tilde{\lambda}}
< 1$) is twice as short as where it is negative (i.e.,  $1< {\tilde{\lambda}}
< 3$), and therefore the mean value may be expected to be negative. And 
indeed, for
an uniform PDF $p({\tilde{\lambda}}|\lambda_3)=1/3$, we recover
$$
\langle \lambda_1\lambda_2\lambda_3|\lambda_3\rangle=
\frac{\lambda_3^3}{4}\int_0^3p({\tilde{\lambda}}|\lambda_3)
(1-{\tilde{\lambda}}^2)d{\tilde{\lambda}}=
-\lambda_3^3/2<0.$$

These considerations appear to be
``too general", and therefore suspicious, because they do not incorporate the 
equation of motion.  And indeed, these
considerations are bias. Recall that $\lambda_3$ (which is $>0$) 
was previously fixed; if we now fix $\lambda_1$($<0$), then
\begin{equation}
\lambda_1\le \lambda_2\le -\frac{\lambda_1}{2},
\label{lambda3}
\end{equation}
cf., (\ref{lambda1}). Calculating now $\langle
\lambda_1\lambda_2\lambda_3|\lambda_1
\rangle =-\langle\lambda_1\lambda_2(\lambda_1+\lambda_2)\rangle$, 
analogously to (\ref{e3}),
we obtain $\langle \lambda_1\lambda_2\lambda_3|\lambda_1\rangle =
-\lambda_1^3/2>0$!

Finally, if we no longer  require (\ref{lambda}), i.e., if we do not
require any ordering for the eigen values, then 
$\langle \lambda_1\lambda_2\lambda_3\rangle=0$. Indeed, all probability
distributions are expected to be symmetric {\it a priori} in respect to the
transformation $\lambda_{1,2,3}\to -\lambda_{1,2,3}$, and therefore 
 $\langle \lambda_1\lambda_2\lambda_3\rangle=0$.
Thus, these ``general considerations" fail to account for the asymmetry,
because they do not incorporate the equation of motion. 
In the framework of K41, the asymmetry {\em is needed} to account for the energy
transfer to small eddies \cite{book}, the asymmetry by itself being expressed 
through the Kolmogorov
law (\ref{K1}). However, the theory does not indicate what  dynamical processes
lead to the asymmetry. 
Another setback in this picture is that the turbulence is
completely non-intermittent, in contradiction with experimental data.

In particular, the asymmetry dictated by the Kolmogorov law, and observed
(negative) skewness, are not supposed to be  related to the intermittency.
That is, the PDF for $\Delta_r u$ is  asymmetric, however,
the asymmetry has nothing to do with the tails of the PDF, that is with 
possible intermittency. 
 In other words, the asymmetry of the PDF is expected to be manifested in the
core. This ``ideal" PDF is constructed in \cite{2000}, in order to compare with
observed properties of turbulence. The PDF, 
\begin{equation}
p(\tilde{u},r)\equiv {\cal{I}}(\tilde{u}),~~~~
\tilde{u}=\frac{\Delta_r u}{\langle (\Delta_r u)^2\rangle^{1/2}}
\label{ideal}
\end{equation}
has the following properties,
$$
\int {\cal{I}}(\tilde{u}) d\tilde{u}=1, ~~~
\int \tilde{u} {\cal{I}}(\tilde{u}) d\tilde{u}=0,~~~
\int \tilde{u}^2 {\cal{I}}(\tilde{u})  d\tilde{u}=1,
$$
and
\begin{equation}
\int \tilde{u}^3 {\cal{I}}(\tilde{u}) d\tilde{u}=
-\frac{4}{5}\frac{1}{C_2^{3/2}},
\label{skewness0}
\end{equation}
where $C_2$ is the Kolmogorov constant. But, most importantly, the PDF is
constructed from two Gaussian functions, so that it does not contain any 
tails. We will use this ``ideal" PDF below; we note for now that the
experimental asymmetry {\em is} related to the intermittency. In particular, 
the tails
of real PDF give a substantial contribution to the Kolmogorov law \cite{2000}.

In addition,
the experimental PDF $p(\tilde{u},r)$ is not self-similar, that is, it is also a 
function of
$r$ (as well as of $\tilde{u}$, cf. (\ref{ideal})), see, e.g., \cite{similarity}.
This in effect implies presence of  intermittency. Direct study of intermittency
can be provided by cumulative moments,
\begin{equation}
S_{0,c}(r)=\int_{-c}^c p(\tilde{u},r)d\tilde{u},
\label{cumulative}
\end{equation}
and
\begin{equation}
S_{-,c}(r)=\int_{-c}^0 p(\tilde{u},r)d\tilde{u}, ~~~
S_{+,c}(r)=\int_0^c p(\tilde{u},r)d\tilde{u},
\label{cumulative+}
\end{equation}
and by the tail moments,
\begin{equation}
S_{0,t}(r)=\int_{-\infty}^{-t} p(\tilde{u},r)d\tilde{u}+
\int_t^{\infty} p(\tilde{u},r)d\tilde{u}=1-S_{0,c=t}(r),
\label{tail}
\end{equation}
and
\begin{equation}
S_{-,t}(r)=\int_{-\infty}^{-t} p(\tilde{u},r)d\tilde{u}, ~~~
S_{+,t}(r)=\int_t^{\infty} p(\tilde{u},r).
\label{tail+}
\end{equation}
If $c$ and $t$ are not small (we will consider $c=t=3$, and $c=t=4$), then 
(\ref{cumulative})
presents major events produced by the core of the PDF, and $S_{0,c}$ should be
close to unity, or rather Gaussian value with which it will be compared. We
denote the
cumulative and tail moments for Gaussian and ``ideal" distributions with
letters $G$ and $\cal I$ correspondingly. For example, $S_{0,c}(r)$ for Gaussian
distribution is $G_{0,c}$, and $S_{-,t}(r)$ for ideal distribution is 
${\cal I}_{-,t}$, etc.
And, of course, $S_{0,c}$ should be independent of $r$, in case of
self-similarity. On the contrary, substantial deviation from unity, and, what is
more important, dependence on the distance -- increasing with growing $r$, 
would indicate intermittency. In fact, super-Gaussian deviation from unity
suggests
that the tails give considerable contribution to the PDF, and this contribution
is decreasing with growing distance. We will see in Sec. ~\ref{4b} 
that that is what is indeed observed.

As to $S_{\pm,c}(r)$, they represent a more subtle way to measure
intermittency, and, of course, asymmetry. These quantities are expected to be
close to (and smaller than) $1/2$, and $S_{+,c}$ being slightly larger than 
$S_{-,c}$.

\subsection{ Extreme intermittency}
\label{2b}

There is a geometrical interpretation of $\langle \lambda_1\lambda_2
\lambda_3\rangle <0$. It implies that typically $\lambda_1<0$, while
$\lambda_{2,3}>0$, that is, the inflow is in one direction, 1, while the outflow
proceeds in two directions, 2,3. Statistically, these flows would generate 
vortex sheets rather than vortex ropes \cite{bet}. However, it became clear 
that the
presence of the vortex itself changes the principal axes and eigen values 
\cite{1,2}. Consider an axisymmetric cell of a scale $\ell$.  At the axis,
there is an inflow in two directions, 1,2, that is, to the axis, and outflow in
one direction, 3, parallel to the axis.
Therefore, locally 
$\lambda_1\lambda_2\lambda_3>0$ ($\lambda_{1,2}<0$ and $\lambda_3>0$).
To be more specific, in cylindrical coordinates $\{r',\phi,z\}$,
\begin{equation}
{\bf v}=\{~~-\alpha r',~~0,~~2\alpha z\},
\label{strain}
\end{equation}
$\alpha =-\lambda_1=-\lambda_2=\lambda_3/2$. Suppose now that there is a
symmetric vortex present, with initial scale $\ell$. Then the vortex is 
 stretched by the strain motion (\ref{strain}), according to equation
\begin{equation}
\partial_t \omega_z-\alpha r'\partial_{r'} \omega_z=2\alpha \omega_z +
\frac{\nu}{r'}\partial_{r'} r'\partial_{r'}\omega_z.
\label{omega}
\end{equation}
This stretching proceeds until viscous effects come into play, resulting in
Burgers vortex,
\begin{equation}
\omega_z(r')=\omega_z(0)e^{-r'^2/\lambda^2},
\label{vortex}
\end{equation}
\onecolumn

\begin{figure}
\psfig{file=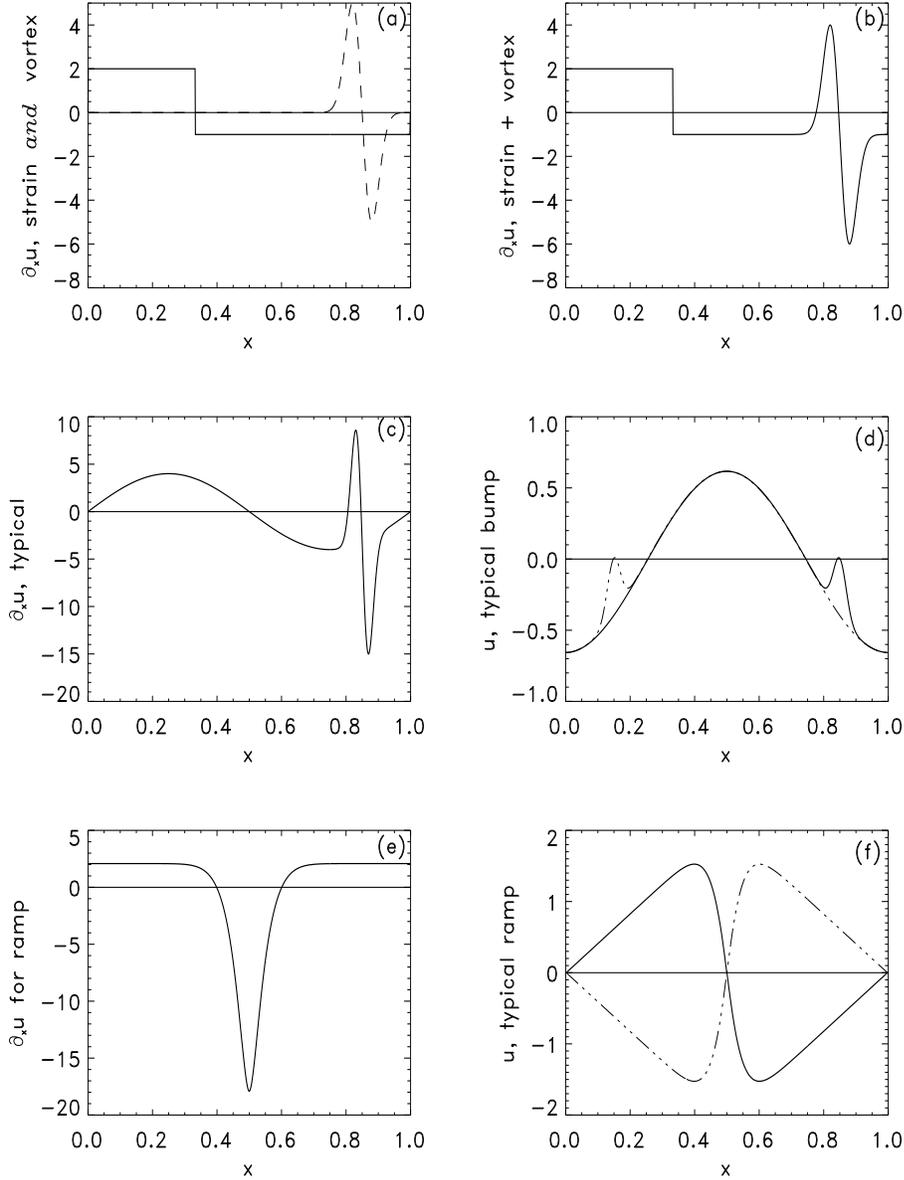,width=5.0in,height=6.5in}
\caption{Schematic presentation of bump model, and its comparison with the
ramp-model. (a) Presentation of $\partial_x u$ for the strain (\ref{strain}),
solid line. The vortex is superimposed on this plot (dashed line). (b) Resulting
motion, the strain plus the vortex. (c) More realistic presentation of the same
as in (b). (d) Corresponding motion $u$. Note typical bump. The dashed-dotted
line presents a bump in a ``wrong place", that is, this situation is
statistically unlikely. (e) $\partial_x u$
for the ramp-model, and (f), corresponding motion, which is indeed ramp-like.
The dashed-dotted line depicts another ramp, which is supposed to appear
statistically rarely. }
\label{f1}
\end{figure}
\twocolumn 
and
\begin{equation}
v_\phi=\frac{\omega_z(0)\lambda^2}{2r'}\left(1-e^{-r'^2/\lambda^2}\right),
\label{vortex1}
\end{equation}
where
\begin{equation}
\lambda=\sqrt\frac{2\nu}{\alpha}.
\label{scale}
\end{equation}
As $\alpha \sim v_\ell/\ell$, $\lambda \sim \ell R^{-1/2}$, and therefore
 $\lambda$ coincides with Taylor microscale (\ref{Taylor1}).
In order to estimate $\omega_z(0)$ we note that, neglecting viscosity, equation
(\ref{omega}) conserves the quantity
\begin{equation}
\int \omega_z r' dr',
\label{conserve}
\end{equation}
and hence we estimate, $\omega_z\sim 1/r'^2$, that is,
\begin{equation}
\omega_z(0)\approx \omega_\ell\frac{\ell^2}{\lambda^2}=\omega_\ell R,
\label{vortex0}
\end{equation}
where $\omega_\ell$ is initial vorticity.

It was shown in \cite{1,2} that the combination of the strain motion (\ref{strain})
and the vortex results in negative skewness. In order to get some simple
estimates, we present $\partial_x u$ corresponding to this motion in Fig.
~\ref{f1}. 
Figure ~\ref{f1}(a) presents the strain (\ref{strain}) with $\alpha=1$. 
The probability
$p(\partial_x u>0)=1/3$ (where $\partial_x u=2$), while $p(\partial_x u<0)=2/3$ 
(and $\partial_x u =-1$), and therefore
$$
\langle \partial_x u_{st}\rangle= 2p(\partial_x u>0)-1p(\partial_x u<0)=0,
$$
and
$$
\langle (\partial_x u_{st})^3\rangle= 8p(\partial_x u>0)-1p(\partial_x u<0)=
8/3-2/3=2>0.
$$
That is, the skewness is positive for the strain motion, as mentioned above.
The vortex corresponds to a sharp odd function concentrated on a small scale
$\lambda$. Therefore, $\langle \partial_x u_\lambda \rangle =0$ and 
$\langle (\partial_x u_\lambda)^3 \rangle =0$. Nevertheless, the sum of these
motions, depicted in Fig. ~\ref{f1}(b), may possess negative skewness.
Indeed,
$$
\langle \partial_x u_{st}+\partial_x u_\lambda \rangle =
\langle \partial_x u_{st}\rangle+\langle\partial_x u_\lambda \rangle =0,$$
while
\begin{equation}
\langle (\partial_x u_{st}+\partial_x u_\lambda)^3 \rangle =
\langle (\partial_x u_{st})^3\rangle-3\cdot 1\langle (\partial_x u_\lambda)^2
\rangle
\label{skewness}
\end{equation}
(cf. \cite{2}),
which becomes negative if the second term on the right-hand-side prevails. 
In particular, for the strain (\ref{strain}) plus vortex (\ref{vortex}),
\begin{equation}
\langle (\partial_x u_{st}+\partial_x u_\lambda)^3 \rangle =
2\alpha^3-3\alpha \omega_\ell^2 R^2\frac{\lambda^2}{\ell^2}\approx -3\alpha
\omega_\ell^2 R,
\label{skewness1}
\end{equation}
where we used the estimate (\ref{vortex0}), and the fact that the vortex
occupies only $\lambda^2/\ell^2$ part of the space (naturally, we also suppose
that $\omega_\ell$ is not small compared with $\alpha$). 

Figure ~\ref{f1}(c) depicts 
essentially the same as Fig. ~\ref{f1}(b), but it suggests a more  realistic
presentation of $\partial_x u$. The integral of this latter function, that is,
the velocity $u$ itself is depicted in Fig. ~\ref{f1}(d). 
Note characteristic bump which
appears due to the vortex. Panels (e) and (f) are to be compared with (c) and
(d); they present $\partial_x u$ and $u$ for the ramp model (the
details of this comparison are given is Secs. \ref{4b} and \ref{5b}).

Imagine now a statistical ensemble of the cells of the scale $\ell$, so that the
ensemble as a whole is statistically isotropic. Suppose that each of the cells
contains a Burgers vortex (\ref{vortex}) with $\omega_\ell\approx \alpha$.
Then,
\begin{equation}
\langle (\partial_x u)^2\rangle \sim \omega_\ell^2 R^2\frac{\lambda^2}{\ell^2}
\sim \omega_\ell^2 R\sim \left(\frac{v_\ell}{\ell}\right)^2 R,
\label{second}
\end{equation}
so that
\begin{equation}
\nu \langle (\partial_x u)^2\rangle \sim \frac{v_\ell^3}{\ell}\sim 
\varepsilon,
\label{energy}
\end{equation}
as in K41 theory. 

 The Kolmogorov law (\ref{K1}) is also
valid, of course. However, the ``inertial range" for this ensemble, starting at
the scale $\ell$ is cut off already
at the Taylor microscale
$\lambda$, rather then  Kolmogorov microscale $\eta$, because the linear in $r$
term in (\ref{K}) is larger than the viscous term only at $r>\lambda$. 
At $r<\lambda$ 
${\cal{S}}_3(r)\sim r^3$, reaching at $r=\lambda$ the amplitude 
${\cal{S}}_3(\lambda) \sim
\varepsilon \lambda$, as should be according to (\ref{K1}). In addition, the
skewness, $\sim {\cal{S}}_3(\lambda)/\lambda^3\sim -\varepsilon / \lambda^2
\approx -(v_\ell/\ell)^3 R$, which coincides with the estimate (\ref{skewness1}). 

So far, this ensemble looks similar to the K41 theory. However the
intermittency in this model is much too high. For example, 
$\langle (\partial_x u)^4\rangle   
\sim \omega_\ell^4 R^4\lambda^2/\ell^2\sim \omega_\ell^4 R^3$, so that the 
flatness,
\begin{equation}
F(0)=\frac{\langle (\partial_x u)^4\rangle}{\langle (\partial_x u)^2\rangle^2}
\sim R,
\label{flatness0}
\end{equation}
which is many orders of magnitude above the experimental value (see, e.g., 
below Sec. ~\ref{4a}). 

\section{ Comparison with experimental data}
\label{3}

Further discussion of these two models and the ramp model is provided below in
Sec. ~\ref{5}. It is clear for now that both extreme cases: no intermittency
~\ref{2a}, and
extreme intermittency ~\ref{2b} are not realistic, and it should be accepted 
something in between.  
Let us accept the K41 theory as a ``first approximation", and we
will try to modify it by introducing finite intermittency. Then,
in addition to what is described in Sec. ~\ref{2a}, 
the cells are generating vortices
as in Sec. ~\ref{2b} above. The question is, what are the initial magnitudes of
$\omega_r$? It is clear that, for example, $\omega_\ell$ should be small
compared with $v_\ell/\ell$: otherwise the intermittency is too strong. Another
possibility is to suggest that 
$\omega_\ell \sim
v_\ell/\ell$, however, only a small fraction of the cells are generating these
vortices.  
In addition, there is nothing special in the large scale eddies; in fact, the
cells with scales $r<\ell$ are generating the vortex ropes as well. Namely, the
scales of generated vortices by the cells of the scale $r$ are,
\begin{equation}
\delta_r =\sqrt{\frac{2\nu}{v_r/r}}=\lambda\left(\frac{r}{\ell}\right)^{1/3},
\label{delta}
\end{equation}
cf. (\ref{scale}). Each cell is able to generate a vortex rope as long as 
$\delta_r <r$, i.e., according to (\ref{delta}), when $r> \ell /R^{3/4}\sim
\eta$, that is, in the whole inertial range. Thus, the vortices are generated in
each cell, however, the scales of these vortices, $\delta_r$, are,
\begin{equation}
\eta \le \delta_r \le \lambda.
\label{intermediate}
\end{equation}

We can call this range intermediate , and it is in this range where the vortices
are indeed observed \cite{vortices}.
\begin{figure}
\psfig{file=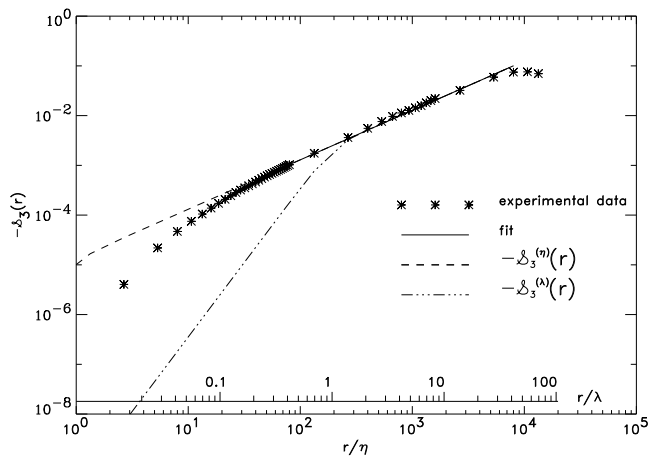,width=3.5in}
\caption{Experimental and theoretical (somewhat simplified and sketchy)
 third moment, ${\cal{S}}_3(r)$. 
\label{f2}
}
\end{figure}

We note that, according to Sec. ~\ref{2b}, the vortex is stretched along the
stretching axis (with maximum value of $\lambda_{1,2,3}$), of course, 
that is, parallel  to the $z$-axis. However,
as mentioned at the beginning of Sec. ~\ref{2b}, when the
vortex is formed, the eigen values are changed. While $\lambda_3$ remains the
same $=2\alpha$, the other two eigen values are now 
$$
\lambda_{1,2}\to  -\alpha \pm \frac{1}{2}\left|\frac{d v_\phi}{d
r'}-\frac{v_\phi}{r'}\right|, $$
where $v_\phi$ is defined in (\ref{vortex1}), see, e.g., \cite{2}.
We now have to arrange the eigen values in order (\ref{lambda}) to obtain
$$
\lambda_1=-\alpha - \frac{1}{2}\left|\frac{d v_\phi}{d
r'}-\frac{v_\phi}{r'}\right|,  ~~~
\lambda_2=2\alpha, ~~~$$
\begin{equation}
\lambda_3=-\alpha + \frac{1}{2}\left|\frac{d v_\phi}{d
r'}-\frac{v_\phi}{r'}\right|.
\label{lambda2}
\end{equation}
We see that now the vortex is aligned along the {\em intermediate} eigenvector.
This feature was initially suggested and shown in \cite{Bob}, and confirmed both
in numerical simulations \cite{vortices}, and in laboratory experiments
\cite{experiment}.

In addition, the eigen values (\ref{lambda2}) result in correct sign of the
skewness (i.e., correct asymmetry of the PDF). Indeed, 
$$\lambda_1 \lambda_2
\lambda_3 = 2\alpha \left[\alpha^2-\frac{1}{4}\left(\frac{d v_\phi}{d
r'}-\frac{v_\phi}{r'}\right)^2\right]<0,
$$
if the amplitude of the vortex is large enough. This formula is consistent with
the estimate (\ref{skewness1}). Indeed, $dv_\phi/dr'-v_\phi/r'\approx
\omega_\ell R^2$, and averaging this expression over the space would result in a
factor $\sim 1/R$, because the vortex is occupying only $1/R$ fraction of space.
As a result, we recover (\ref{skewness1}).

\subsection{The third moment}
\label{3a}

In order to understand what impact has this model on  the
structure functions, we start with the third order structure function. As
mentioned, for the extreme intermittency case, Sec. ~\ref{2b}, ${\cal S}_3 \sim
r^3$ for $r<\lambda$, while for no-intermittency case, Sec. ~\ref{2a}, 
${\cal S}_3 \sim r^3$ is valid for $r<\eta$. Naturally, the picture outlined
above in this section suggests intermediate situation. 

Figure ~\ref{f2} depicts experimental ${\cal S}_3(r)$, and it is
compared with what we would expect for the ensemble described in Sec. 
~\ref{2b}, namely,  with the following function  constructed for this purpose
$$
{\cal S}_3^{(\lambda)}(r)=-\frac{4}{5}r\varepsilon \tanh{\left(\frac{r}{\lambda}
\right)^2}.
$$
 The experimental ${\cal{S}}_3(r)$ is also compared with
$$
{\cal S}_3^{(\eta)}(r)=-\frac{4}{5}r\varepsilon \tanh{\left(\frac{r}{\eta}
\right)^2},
$$ 
which is  what one would expect from K41. Note that all three curves,
one experimental and two constructed, match at inertial range, i.e., at $r>
\lambda$. In order to do this, the constructed functions are matching
experimental linear fit with exponent $\zeta_3=0.99\pm 0.01$, quite close to
the Kolmogorov law (for which $\zeta_3=1$). The plots are given in Kolmogorov
microscale, that is, in $r/\eta$, and there is also $r/\lambda$ scale in the
figure for comparison. It can be seen that the inertial range indeed  starts at
$r\approx \lambda$, and thus the intermediate range (\ref{intermediate})
is clearly visible. Besides, the experimental plot is situated between no
intermittency moment, ${\cal{S}}_3^{(\eta)}(r)$, and extreme intermittency, 
${\cal{S}}_3^{(\lambda)}(r)$, -- as expected.
\onecolumn

\begin{figure}
\psfig{file=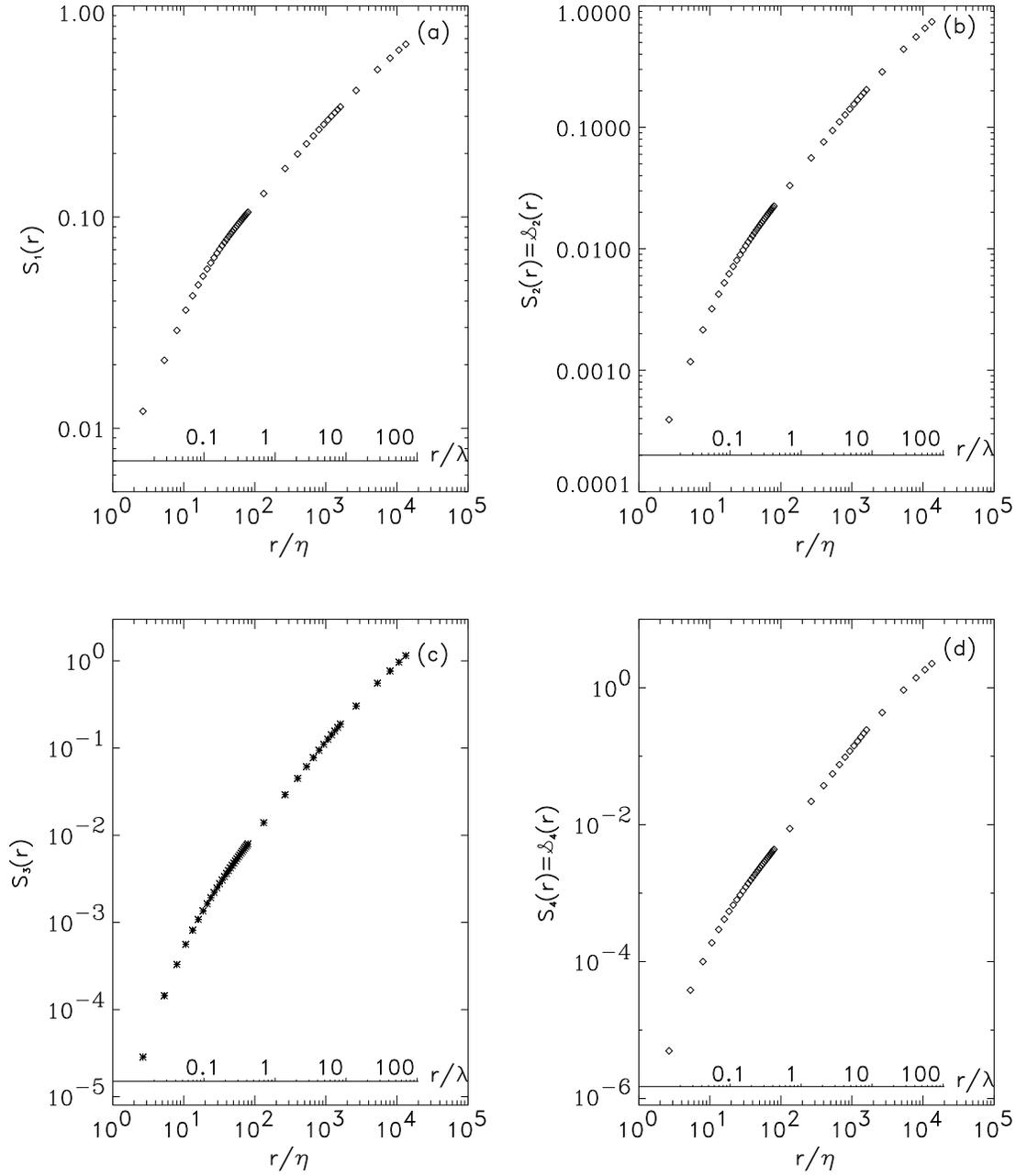,width=6.0in,height=7in}
\caption{Panels (a) through (d): the generalized structure functions: 
starting from the first and ending with the fourth.
\label{f3}
}
\end{figure}

\begin{figure}
\psfig{file=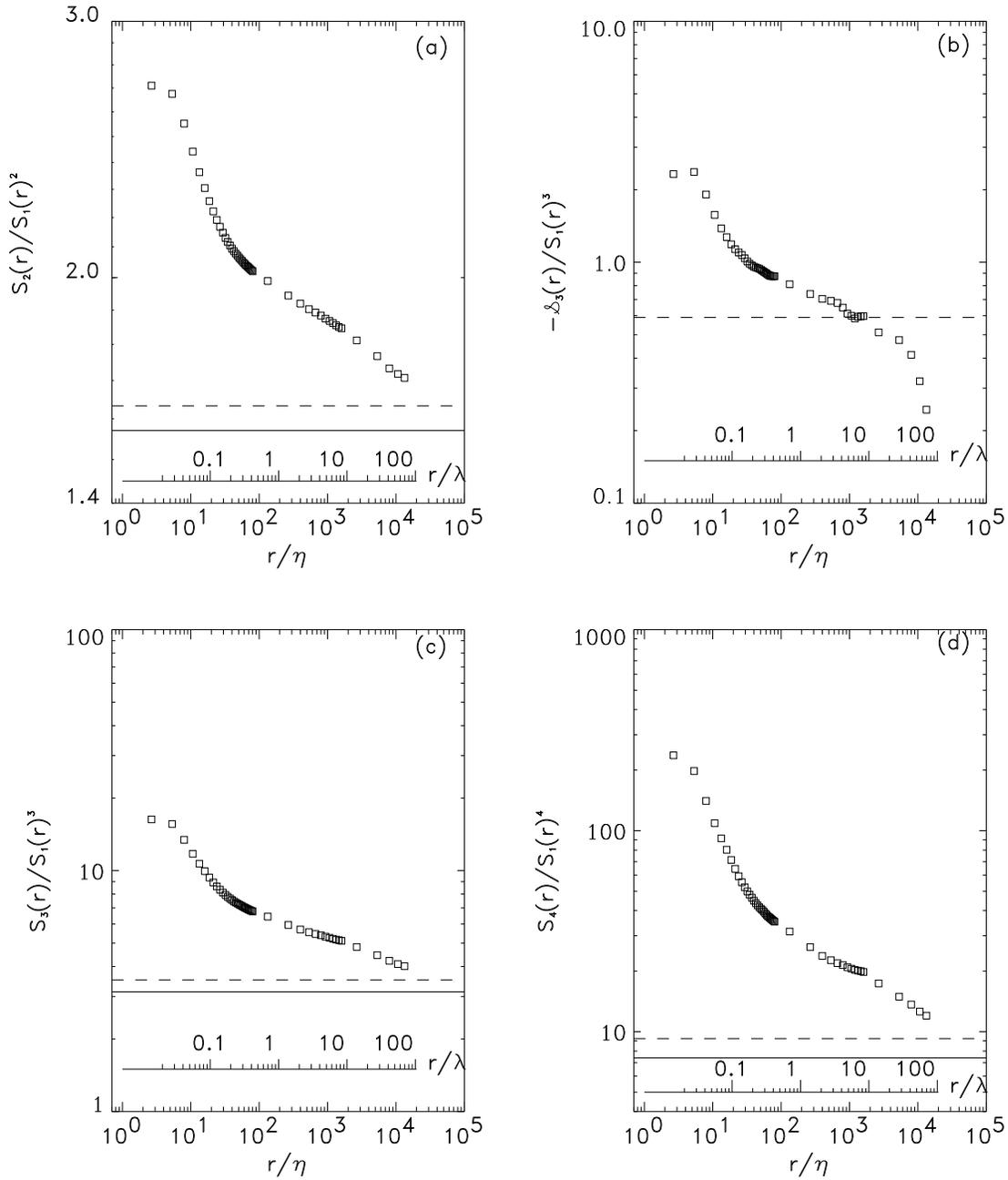,width=6.0in,height=7in}
\caption{Generalized flatness; several dimensionless ratios of different moments
to the first. The plots are compared with the Gaussian value (solid line), and
with what would follow from the ``ideal" PDF, see (\ref{ideal}-\ref{skewness0}).
Note that all plots are in log-log scale. The ordinate in panel (a) is in
log-scale as well; it seems indistinguishable from linear scale because the
scale range is relatively small.
\label{f4}  
}
\end{figure}

\begin{figure}
\psfig{file=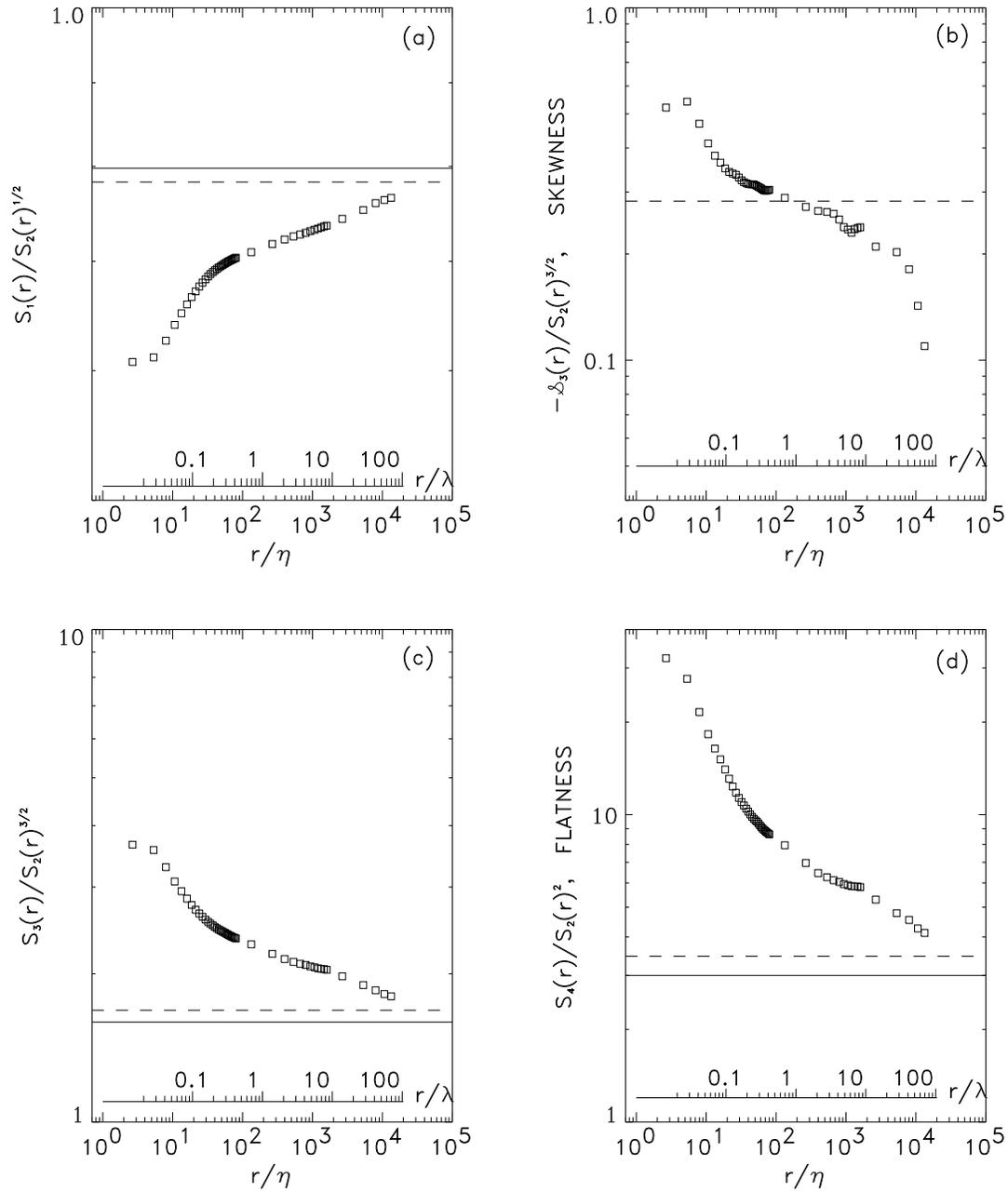,width=6.0in,height=7in}
\caption{Same as in previous figure, only these are ratios of moments to the 
second order structure function.
\label{f5}
}
\end{figure}
\twocolumn

\subsection{Other moments}
\label{3b}

The inertial range is associated with scaling, that is, in the log-log plotting
the moments are presented with straight lines. 
Figure ~\ref{f3} illustrates generalized structure functions of different orders, 
the first, the second, the third and the fourth (and, of course, the 
structure functions of even orders coincide with the generalized  structure 
functions). 
The intermediate range (\ref{intermediate}) is easily separated  from the
inertial range (where the plots become straight lines).

We may interpret this separation of the intermediate range by the fact that 
this range contains vortices. This interpretation is not unequivocal however.
It is also possible that the depletion of the structure functions in this
range, as compared with the K41, is due to the action of viscosity. Indeed, the
viscosity is still acting even at $r\gg\eta$, in spite of the fact that the local
Reynolds number is there $\gg 1$. The fact that the viscous effect is stretched
up to $r\sim \lambda$ might be  purely coincidental. 

\section{Intermittency}
\label{4}

\subsection{Flatness}
\label{4a}

A real test to check if the intermediate range is related to vortices is 
achieved by the measuring the
intermittency. Traditionally, the later is estimated by measuring the
flatness, see, e.g., \cite{book}. More specifically, the flatness, and related
quantities, like generalized flatness, are compared with corresponding 
Gaussian values. Figure ~\ref{f4} presents
dimensionless ratios of different moments to
the first moment. They can be called generalized flatness. We see that, first, 
the intermediate range is clearly distinguished from inertial, the intermittency
growing substantially in the intermediate range. The deviation from Gaussian and
``ideal'' behavior is especially dramatic in Fig. ~\ref{f4}(d), which is hardly
surprising taking into account that it represents a relatively high moment.
 
Second, although the
deviation from Gaussian and ``ideal" values in inertial range is not big, (a
factor of 2 or 3), there is a noticeable scaling for about two decades. Recall
that in the framework of K41, according to (\ref{K41}), all these ratios would
be flat, that is, e.g., $S_2(r)/S_1(r)^2= \rm const$, etc. If however the 
structure
functions have anomalous scaling, $S_q(r)\sim r^{\zeta_q}$, where $\zeta_q \not=
q/3$, then these ratios would possess some scaling, which is indeed observed.

Finally, what may be called  generalized skewness, depicted in Fig.
~\ref{f4}(b), is
presented by rather scattered curve. This might be explained as follow. 
The third moment of the structure function, appearing only due to the asymmetry
(as opposed to the third order generalized structure function, depicted in panel
(c)),  contains quite a substantial contribution coming from the PDF tails, 
see, e.g., \cite{98}. As the tails correspond to rare (but stormy) events, the
odd order structure functions are subject to substantial fluctuations. 

Essentially the same conclusions can be drawn from analyzing Fig. ~\ref{f5}, 
where the
dimensionless ratios of different moments to the second order structure function
are depicted. It is clear that the intermittency is substantially increased in
the intermediate range. In particular, for the ratio $S_1(r)/S_2(r)^{1/2}$, the
presence of intermittency implies that the ratio is {\it below} the Gaussian
value, see Fig. ~\ref{f5}(a), rather than otherwise for higher moments; 
and still, 
the intermediate
range is clearly distinguished by enhanced intermittency. It is also clear from
panel (b) that the skewness substantially exceeds that given by the ``ideal" PDF
only in the intermediate range. In the inertial range, the skewness is
decreasing systematically with increasing distance $r$, although the deviation
from ``ideal" behavior is not conspicuous (and the scaling, if any, is rather
poor). Finally, maximal flatness (panel (d)) reaches the value $32.28$, well
below what would be expected from extreme intermittency model described in Sec.
~\ref{2b}. 

\subsection{Box counting}
\label{4b}

Zeroth moment of structure function, that is box counting, are very
useful to study because, unlike high moments, the statistic is as good as one
can get, although quite often the zeroth moment provides only trivial results. 
In order to extract some useful data in studying intermittency, one can either
analyze the cumulative (and tail) moments, or separate negative values from the
positive, and to study them separately (as, e.g.,  in \cite{2000}). 

\begin{figure}
\psfig{file=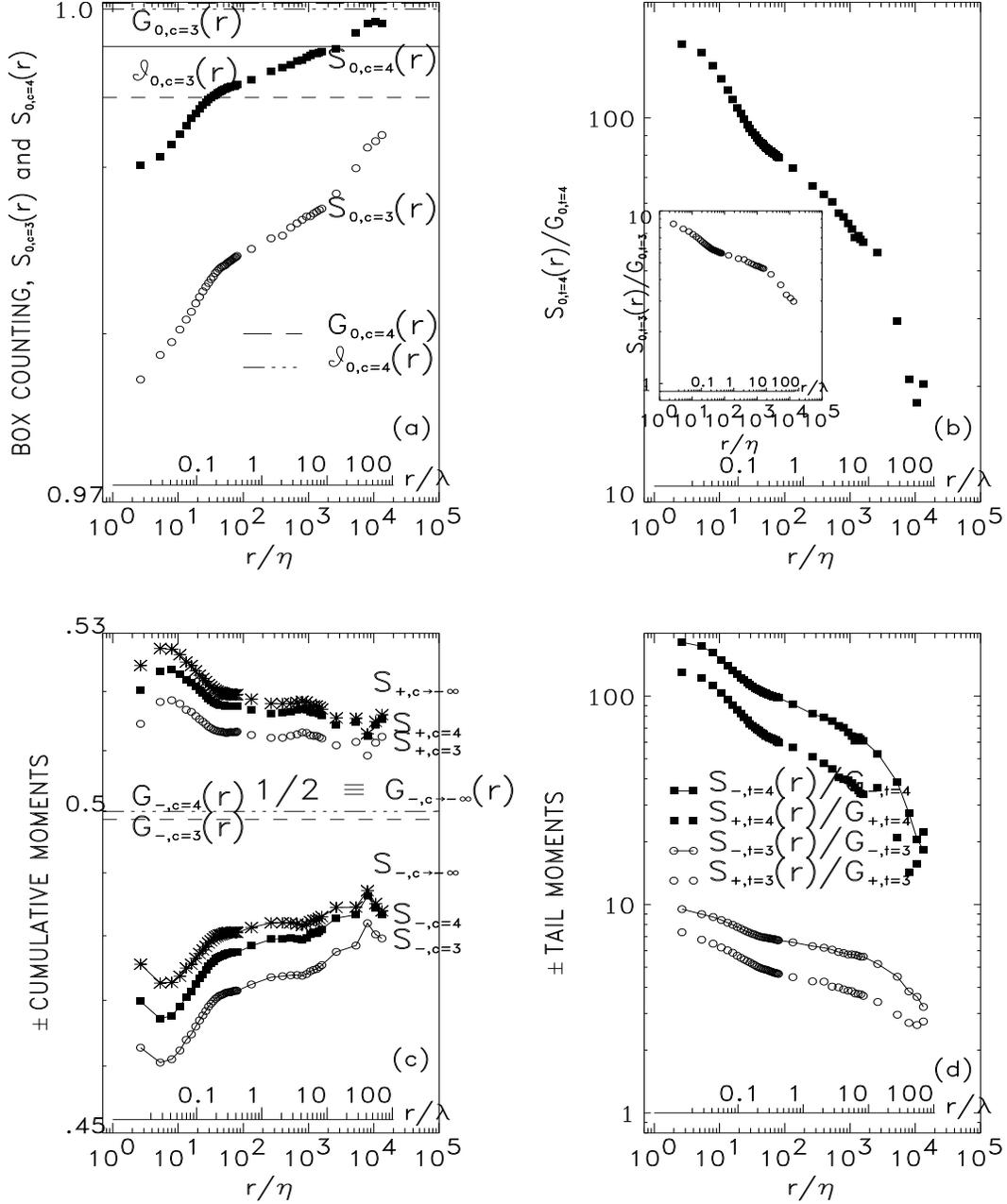,width=6.0in,height=7.0in}
\caption{Cumulative and tail moments of zeroth order. (a) The  probability of
$|\tilde{u}|\le 3$, and of $|\tilde{u}|\le 4$ for different $r$'s, which coincide 
with $S_{0,c=3}(r)$ and $S_{0,c=4}(r)$ correspondingly. These quantities are
compared with correspondent Gaussian and ``ideal" values, $G_{0,c=4}$ being
almost indistinguishable from unity.
(b) Tail moments, with respect to correspondent Gaussian values, and for the
same values $t=3$ and $t=4$, are depicted. (c) The probability of
$\tilde{u}<0$, corresponding to $S_{-,c\to -\infty}(r)$, probability of
$-3 \le \tilde{u} \le 0$, or $S_{-,c=3}(r)$, and of $-4 \le \tilde{u} \le 0$, or
$S_{-,c=4}(r)$ are depicted, and compared with corresponding Gaussian values.
This time, $G_{-,c=4}$ is practically indistinguishable from $1/2$.
(d) The probability of $\tilde{u} \le -3$ (or $S_{-,t=3}(r)$), and of $\tilde{u} 
\le -4$ (or $S_{-,t=4}(r)$) are depicted in respect to their Gaussian values.
\label{f6}
}
\end{figure}

Panel (a) and (b) in Fig. ~\ref{f6} correspond to  experimental cumulative and 
tail moments, defined in (\ref{cumulative}) and (\ref{cumulative+})
correspondingly, and we compare them with theoretical values. As mentioned in
Sec. ~\ref{2a}, for self-similar PDF's, the cumulative moments are independent
of distance $r$, and they are expected not to deviate much neither from Gaussian
nor from ``ideal" asymmetric distribution, - these two happen to be close to
each other, as seen from panel (a). On the other hand, the contribution of the
tails, that is, intermittency, if present, is increasing with decreasing 
distance. That
implies that the cumulative moments, that is, the cores of the PDF's give a 
smaller
contribution for decreasing distances. Figure ~\ref{f6}(a) shows that this is 
indeed
the case. It can be seen that, first, the cumulative moments noticeably deviate
from both Gaussian and ``ideal" values. Second, this deviation is higher in the
intermediate range, and that is the way this range is easily separated from
the inertial range.  One can see analogous trends from panel (b),
where tail moments are depicted. The tails definitely give larger contribution
for small distances, and this is especially true for the intermediate range,
which is again noticeable. Note also that the tail parts for $t=4$ are
substantially larger than Gaussian values, especially for
\onecolumn

\twocolumn
\noindent
the intermediate
range. 

We now proceed to the analysis of the $\pm$ moments. It is clear that there are
several common features for ramp model and bump model, that is, common
features in panels (c) and (e) of Fig. ~\ref{f1}. As mentioned in Sec. 
~\ref{2b}, for
both cases, $\langle \partial_x u \rangle=0$ and  $\langle (\partial_x u)^3 
\rangle <0$. This actually implies that the positive part of the distribution
occupies larger space than the negative part. This is seen directly from panel
(e), and can be calculated for the function depicted in panel (c), i.e.
calculated
directly from the function $\partial_x u_{st}+\partial_x u_\lambda$ defined in
Sec. ~\ref{2b}. In other words, $S_{-,c\to \infty}(r)$ should be less than 
$S_{+,c\to \infty}(r)$. As, on the other hand, $S_{-,c\to \infty}(r)+
S_{+,c\to \infty}(r)=1$, by definition, we have 
$S_{-,c\to \infty}(r)<1/2$, and this quantity is typically
increasing with growing $r$, while $S_{+,c\to \infty}(r)>1/2$, decreasing with $r$.
This is exactly what is observed in panel (c) of Fig. ~\ref{f6}, although 
the curves are
somewhat scattered. In spite of the latter, there is no doubt that $S_{-,c}(r)<
S_{+,c}(r)$, for all distances and all values of $c$. 

The box counting $S_{\pm,c=3}(r)$ and $S_{\pm,c=4}(r)$
behaves analogously. The intermediate range can be again distinguished in all of
the curves depicted in panel (c), although the range is not that pronounced.
This range can be better noticed from the tail moments, depicted in panel (d). 
It is clearly
seen that the tails are stronger at small distances, and the intermediate 
range  clearly noticeable. Besides, the negative distribution is definitely more
singular than the positive, or $S_{+,t=3,4}(r)<S_{-,t=3,4}(r)$: and that is 
what is
seen from this panel. This difference in singularity strength is obvious from
the ramp model, Fig. ~\ref{f1}(e), where the positive distribution is not 
singular at
all. It is also seen from Fig. ~\ref{f1}(c), that is, for the bump model, that 
the
negative peak reaches higher absolute 
values than the positive peak, and
therefore the negative singularity is expected to possess higher strength.
Finally, again as in panel (b), the tails deviate quite substantially from
Gaussian values for $t=4$.

\section{Discussion of the model}
\label{5}

\subsection{Intermittency formation}
\label{5b}

Let us return to the stretching process described in Sec. ~\ref{2a}. In more
general case of axisymmetric stretching we have,
\begin{equation}
\frac{\partial v_\phi}{\partial t}
+\left[v_{r'}\frac{1}{r'}\frac{\partial  
r'}{\partial r'}+v_z\frac{\partial}{\partial z}
\right]v_\phi=\nu \left[\frac{\partial^2}{\partial z^2} +
\frac{\partial}{\partial r'} \frac{1}{r'}
\frac{\partial}{\partial r'} r'\right]v_\phi.
\label{omega1}
\end{equation}
In the vicinity of stagnation line $r'=0$, this equation is reduced to
(\ref{omega}) if operator $\nabla\times$ acts on (\ref{omega1}). This equation
is {\em linear} in respect to $v_\phi$, and, if $v_{r'}$ and $v_z$ are given,
then the vortex stretching, described by equation (\ref{omega1}) can be
considered {\em kinematic}. In fact, this equation coincides with that for
$A_\phi$, 
$\phi$-component of magnetic vector-potential, i.e., $B_{r'}=-\partial_z A_\phi$,
$B_z=r'^{-1}\partial_{r'} r' A_\phi$, and $\bf B$ is magnetic field strength. 
The
induction equation for $A_\phi$, coinciding with (\ref{omega1}), is kinematic if
the velocity is given. 

The equations for $r$ and $z$-components of the  velocity field read,
$$
\frac{\partial v_{r'}}{\partial t}+
\left[v_{r'}\frac{\partial  
}{\partial r'}+v_z\frac{\partial}{\partial z}
\right]v_{r'}-\frac{v_\phi^2}{r'}=$$
\begin{equation}
-\frac{1}{{\rho}}\frac{\partial p}
{\partial r'} 
+\nu \left[\frac{\partial^2}{\partial z^2} +
\frac{\partial}{\partial r'} \frac{1}{r'}
\frac{\partial}{\partial r'} r'\right]v_{r'} +F_{r'},
\label{momentum}
\end{equation}
and
$$
\frac{\partial v_z}{\partial t}+
\left[v_{r'}\frac{\partial  
}{\partial r'}+v_z\frac{\partial}{\partial z}
\right]v_z=$$
\begin{equation}
-\frac{1}{\rho}\frac{\partial p}{\partial z} 
+\nu \left[\frac{\partial^2}{\partial z^2} +
\frac{\partial}{\partial r'} \frac{1}{r'}
\frac{\partial}{\partial r'} r'\right]v_z+F_z,
\label{momentum1}
\end{equation}
where ${\rho}$ is density, and $\bf F$ is external forcing.

In the vicinity of stagnation line $r'\to 0$ where there is the strain
(\ref{strain}) motion plus the vortex, all the terms in equation
(\ref{momentum}) are functions of $r'$ only, and therefore they can be
compensated by the pressure. Analogously, all the terms in (\ref{momentum1}) are
functions of $z$ only, and again are compensated by the pressure. This means the
``the cell", i.e., the $r$ and $z$-components of the velocity can be considered
as given (and steady). In general, however, $v_\phi$ depends also on $z$, and
therefore the third term in (\ref{momentum}) gives nontrivial contribution to 
the equation. Nevertheless, as mentioned at the beginning of Sec. ~\ref{3},
the velocity corresponding to the vortex is supposed to be small compared with
the strain motion. In other words, the contribution of this term can be
neglected. Besides, the external forcing is always able to make the cell steady.
In any rate, if this term is neglected, then these two equations
(\ref{momentum}-\ref{momentum1}) are independent from existing vortex, and, 
from
the point of view of the vortex generation, the $v_{r'}$ and $v_z$-components 
can be considered as given.

In this -- somewhat simplified -- picture, the vortex generation responsible for
appearance of the intermittency, is kinematic,
that is, passive. We may expect therefore that there is some analogy between this
generation of intermittency and that of a passive scalar gradient. It is known
that the latter is shown to be intermittent even if the background, that is, the
given velocity is not \cite{scalar}. 

Generally, we have the following energy balance equation,
\begin{equation}
\frac{d}{dt} \frac{\langle v^2+v_\omega^2\rangle}{2}=-\nu \langle
({\bf \nabla\times 
 v})^2+({\bf \nabla\times v_\omega})^2\rangle,
\label{energy1}
\end{equation}
where $v_\omega$ stands for the generated vortex. The first term on the
right-hand-side corresponds to regular Kolmogorov cascade, and corresponding
energy dissipation, and the second one -- to the dissipation of the vortices.
In the no intermittency model, Sec. ~\ref{2a}, the second term  is absent.
On the contrary, in the extreme intermittency model Sec. ~\ref{2b}, the first
term can be neglected, because of the large size of the cells, while the second
makes up for the Kolmogorov dissipation if the initial amplitude of the vortex
$\omega_\ell$ is comparable with $\alpha_\ell$, see (\ref{energy}). In a more
realistic model described at the beginning of Sec. ~\ref{3}, both dissipation
terms
contribute. More specifically, the Kolmogorov cascade prevails, and the
generated vortices give a small contribution to the dissipation (otherwise the
intermittency is too strong). Let $\omega_\ell \approx f\alpha_\ell$, 
where $f<1$, then 
\begin{equation}
\nu \langle ({\bf \nabla\times v_\omega})^2\rangle \sim f^2 \frac{v_\ell^3}
{\ell} \sim f^2\varepsilon,
\label{energy2}
\end{equation}
cf. (\ref{second}). It is because of the smallness of $f^2$ that the process of
vortex generation can be considered  kinematic.

Presumably, the most vulnerable part of this picture is that the parameter $f$
is essentially arbitrary. We can suggest some speculation about the origin of 
this
parameter. Namely, the process of vortex formation, and vortex existence, are
transitional. Indeed, the vortex cannot exist longer than the cell itself, 
where the vortex is generated. According to K41, a cell of the size $\ell$ exists
only for a time $\sim \ell/v_\ell$. On the other hand, $1/\alpha_\ell \sim 
\ell/v_\ell$ as well. In other words, the vortex generation time,
$1/\alpha_\ell$, and the life time of the cell, and therefore the life-time
of the vortex
itself, are comparable. Of course, these are just rough estimations. There is
some dispersion both in the life-times, and in the generation times. We may
expect, that a cell should be persistent, in order to have time enough to
generate a vortex. That is to say that only a fraction of cells (which are
persistent) would give rise to vortices. Another restriction arises from
geometry. We may expect that there are vortex perturbations in each cell.
However, as described in Sec. ~\ref{2b}, a cell would generate a vortex only if
the initial perturbation is axisymmetric, with the axis parallel to the
stagnation line of the cell. Generally, a perturbation, even with $\omega_\ell
\approx \alpha$, would not be exactly the way described in Sec. ~\ref{2b}.
Namely, it would not be expected to be axisymmetric at all. Considering again
the interaction to be kinematic, we may expect that only axisymmetric part of
the initial perturbation, -- being generally a small fraction of the
perturbation as a whole, -- is transformed into a vortex, considered in Sec.
~\ref{2a}. Another solution to this problem can be seen in the fact that the
generation of the vortices does not necessary proceeds in the largest cells. Of
course, flatness according to (\ref{flatness0}) $\sim R$ is much too high. If,
however, the generation proceeds from a cell of a size $r$, then, the flatness
would be 
$$F(0) \sim R_r=R\left(\frac{r}{\ell}\right)^{4/3}<R.$$ 

Another difficulty in this model is that it predicts the presence of
intermittency only in the intermediate range, (\ref{intermediate}) in Sec.
~\ref{3}. However, we can see from practically all the plots that the
intermittency is present in the inertial range as well, although, being 
less intense. It can be seen, for example, from Figs. ~\ref{f4} and ~\ref{f5},
where flatness is depicted, that the latter has a decent scaling in the
inertial range (for two decades!). Nontrivial behavior in inertial range can be
also observed in Fig. ~\ref{f6}, especially in panels (a) and (b). 
The tail events are
particularly noticeable for the tails with $t=4$, and their deviation from
Gaussian is substantial in numbers, as seen from panel (b), although these
events are less pronounced than in intermediate range (and the curve is more
scattered). 

We may suggest the following explanations. First, as noted above in this
section, the vortices are transitional events, being generated at large scales,
and then, with diminishing scales they reach their final scale $\lambda$. Thus,
at some point in time, each vortex has a size comparable with $r$, and therefore
it may contribute to the distribution. Second, 
strictly speaking, the structures of smaller (and larger) than $r$  scales do 
contribute to the structure functions taken  at a distance $r$. Therefore
one may expect that the structure functions for distances $r$ corresponding 
to the  inertial range 
would reflect the intermittency of smaller scales, or possibly the
intermittency of the intermediate range only. 

\subsection{Ramp model versus bump model}
\label{5a}

As we noticed in previous section, there are several common features between
these two models. Here, we focus on differences. 
First, the estimates (\ref{skewness}-\ref{skewness1}) differ for
the ramp model. The first moment is still zero, of course, $\langle \partial_x
u_{ramp}
\rangle=\omega\cdot (r-\delta)- [(r-\delta)/\delta] \omega \cdot\delta=0$.   
Here $\delta$ is
the size occupied by  the negative part, and therefore $r-\delta$ corresponds 
to the
remaining positive part, the amplitude of the positive part being
$\omega$, see Fig. ~\ref{f1}(e). The third moment is negative. Indeed,
\begin{equation}
\langle (\partial_x u_{ramp})^3\rangle=\omega^3 (r-\delta) - \left(
\frac{r-\delta}{\delta}\omega\right)^3\delta<0
\label{ramp}
\end{equation}
which is negative if $\delta<r$. Therefore, the skewness is negative as in  
(\ref{skewness}-\ref{skewness1}), but the parameters entering this expression
are different.

The main conclusion about the ramp model is that the strength of the negative
singularity is higher than that of the positive,
\begin{equation}
D_q^-<D_q^+,
\label{dimensions}
\end{equation}
where $D_q^\pm$ are generalized dimensions, \cite{94}. In the simplified picture
illustrated in Fig. ~\ref{f1}(e), the positive part is not singular at all, 
that is,
$D_q^+\equiv 1$. On the other hand, the bump model does suggest that the
positive part is singular, although the inequality (\ref{dimensions}) is 
satisfied:
it is seen from Fig. ~\ref{f1}(c), and follows from discussion at the end of 
previous
section. The singularity of positive tails is indeed observed, see Fig.
~\ref{f6}(d),
and that may be considered as a preference of the bump model as compared with
the ramp model. 

Generally, the bump model can be considered as a more subtle and sophisticated
version of the ramp model, the former suggesting essentially the same 
predictions as the latter.  Obviously, the ramp model is
purely empiric, while the bump model describes some dynamical processes in
formation of vortices. Indeed, Fig. ~\ref{f1}(f) presents two ramps, one is 
what is
expected, and another one (dashed-dotted line) corresponds to a statistically
unlikely ramp. This statistical preference is so far not understood. It was
suggested in \cite{94} that the situation is analogous to a shock wave
formation: a shock would never develop in the shape like depicted with 
a dashed-dotted line, and it evolves only into ``real ramp", (solid line). This
analogy is of little help, however, the typical turbulence being incompressible,
and the shocks are therefore irrelevant. On the other hand, the bump should 
appear on
descending part of the curve, where the derivative is negative, as depicted in
Fig. ~\ref{f1}(d), solid line, as opposed to ``unrealistic" bump depicted with 
dashed-dotted line. This statistical preference can be explained, as opposed to
the ramp-model statistics. Indeed, the  ascending part of the curve (with
positive derivative) corresponds to a structure function with vector $\bf r$
parallel to the vertical direction, where the derivative ($=2\alpha$) is indeed
positive, while the descending part corresponds to $\bf r$ in horizontal plane
(where the derivative is $=-\alpha<0$). Now, the bump corresponds to the vortex,
and, of course, the corresponding velocity is suited in the horizontal plane.
Therefore, the bump would appear only on descending part of the curve. 

Finally, let us discuss the quantitative estimates for the asymmetry. As
mentioned in Sec. ~\ref{2a}, the non-intermittency scenario does not account for
the asymmetry. The latter is easily explained in extreme intermittency approach,
see Sec. ~\ref{2b}. The corresponding expression (\ref{skewness1}) does provide 
a realistic estimate of the skewness, as noted at the end of ~\ref{2b}.
However, this estimate corresponds to extreme intermittency scenario, i.e., to 
{\em unrealistic intermittency}. Still, the asymmetry  can be explained if we
suggest that the vortex formation proceeds all the time. That is to say that the
transitional stage mentioned in previous subsection is always present. If there
is enough time for the perturbation to reach the Taylor microscale, then the
vortex is complete, and therefore the intermittency is generated. But,
considering the fact that the intermittency is not that pronounced in fully
developed turbulence, we conclude that the final formation of a vortex is a
relatively rare event. More often, the vorticity of a perturbation is enhanced
for a while, and then the generating cell ceased to exist, and thus the
generation stops. However, during this amplification, the vorticity of the
generated vortex easily exceeds the causing this generation strain, and then,
according to (\ref{skewness}), the third order structure function becomes
negative. This is valid for any particular scale $r$, if the vortex reaches it
(and its vorticity exceeds the strain). This may account for the Kolmogorov law.

\section{Conclusion}
\label{6}

It is known that production of vorticity, or vortex stretching, does not 
necessary
implies that intermittency is generated. Indeed, in classical Kolmogorov picture
K41, the energy cascade from large eddies to  small actually implies that
small scale vortices are generated: because the vorticity production proceeds
homogeneously. In other words, each eddy decays into smaller eddies, {\em each
of which} in turn decays into smaller eddies, and so on. 

As we argued above, the vortex stretching may be considered as kinematic, in some
limited sense, of course. For a passive scalar $\psi$, the classical Kolmogorov
picture is not much different from what is described in the previous paragraph.
That is, $\psi^2$ is transported down the scales, so that $\langle (\nabla
\psi)^2 \rangle$ is growing (which is similar to the vorticity production).
The energy cascade is gradual and continuous in this picture, meaning that the
large scale fluctuations generate smaller scale perturbations, which in turn
generate even smaller scales, and so on, until the perturbations reach
diffusive scale where they eventually disappear.  We can imagine a {\em direct}
cascade of energy, without passing the intermediate scales in inertial range,
directly from the large scales to the dissipation scale. To be more specific
suppose, that the velocity turbulence is presented by large scale cells only
(of the size $\ell$):
like in Sec. ~\ref{2b}, but the velocity field is random in time, unlike
time-independent, i.e., steady cells from Sec. ~\ref{2b}. Generally, maxima of $
\nabla \psi$ will be generated in  the vicinity of the stagnation lines and
stagnation surfaces,
in a time-scale $\sim \ell/v_\ell$. The $\nabla \psi$-field will have the 
scale of Taylor
microscale. However, if the life-time of the eddies $\tau_\ell\approx \ell/
v_\ell$, as in regular turbulence, then, first, there would not be enough time
to produce a sharp and concentrated structure of the $\nabla \psi$-field during
one turn-over time, and second, in the next life-time, the concentrated 
$\nabla \psi$-field structure is formed in a different place. As a result, the
$\nabla \psi$-field is generated homogeneously in statistical sense. In other
words, there is no intermittency generated. Suppose now that the turbulent cells
are persistent, or, at least, some of them are. Then, distinct $\nabla 
\psi$-field structures are generated, and they are as persistent as the cells
themselves. The scenario described in Sec. ~\ref{2b} of vortex generation 
is essentially the same.

We presented an evidence of the presence of vortices in high Reynolds
turbulence. This is clear from the figures: the intermediate scale range
(\ref{intermediate}), where these vortices are supposed to appear (see the
beginning of Sec. ~\ref{3}, and also \cite{vortices}), is clearly noticeable. 
The main feature which  distinguishes
this range is a substantially enhanced intermittency. The vortices are aligned to
the intermediate eigen-vectors, \cite{Bob}, \cite{vortices}, \cite{experiment}.
And, most important, these vortices account for negative skewness, that is, they
result in a right asymmetry of the PDF. This picture promoted a development of the
ramp model  which is modified now into the bump model. The latter not only
explains more features than the former, but also the bump
model seems to be dynamically substantiated, unlike the ramp model, see Sec. 
~\ref{5a}. 

The most important conclusion is that the intermittency is related to the
asymmetry of statistical properties of turbulence, which was suggested in
\cite{94}. This give us a powerful tool for studying the intermittency: because
the asymmetry is manifested in lower moments of structure function, while the
intermittency normally manifests itself in higher moments with poor statistics.

Still, the whole picture is far from complete. As mentioned in Sec. ~\ref{3},
this simple scenario, outlined again above in this section, predicts too much
intermittency. There are some suggestions in Sec. ~\ref{3} as to how to cure this
difficulty. But, so far, the theory is unable to predict {\em the numbers}, e.g.,
we cannot predict the flatness from the asymmetry, or the other way.
Nevertheless, we believe that the paper provides a next step in understanding
the intermittency of turbulence.

\acknowledgments
I thank K. R. Sreenivasan and B. Dhruva for   the data of high Reynolds number
atmospheric turbulence used in this paper. I also appreciate discussions with 
W. Goldburg, J. Gollub, X. L. Wu, and Z. Warhaft.

\end{document}